\documentclass[letterpaper, 10 pt, conference]{ieeeconf}  % Comment this line out if you need a4paper

\IEEEoverridecommandlockouts                              % This command is only needed if 
\usepackage[utf8]{inputenc}
\usepackage{graphicx}
\usepackage{amsmath}
\usepackage{hyperref}
\usepackage{draftwatermark}

\usepackage[markup=bfit, deletedmarkup=sout, authormarkup=brackets]{changes}
\definechangesauthor[name={Hagen}, color=teal]{HL}
\definechangesauthor[name={Matej}, color=orange]{MH}
\definechangesauthor[name={Adam}, color=red]{AR}
\definecolor{Gray}{gray}{0.9}
\definecolor{Gray2}{rgb}{0.1,0.1,0.1}

\title{\LARGE \bf Should a small robot have a small personal space? Investigating personal spatial zones and proxemic behavior in human-robot interaction}
\author{Hagen Lehmann$^{1, 2}$ and Adam Rojik$^{1}$ and Matej Hoffmann$^{1}$ % <-this % stops a space
\thanks{M.H. and A.R. were supported by the Czech Science Foundation (GA {\v C}R), project EXPRO (nr. 20-24186X). H.L. was supported by the International Mobility of Researchers in CTU, 
Nr. CZ.02.2.69\/0.0\/0.0\/16\_027\/0008465. }% <-this % stops a space
\thanks{$^{1}$The authors are with the Department of Cybernetics, Faculty of Electrical Engineering, Czech Technical University in Prague, Czech Republic.{\tt\small matej.hoffmann@fel.cvut.cz}}%
\thanks{$^{2}$Hagen Lehmann is with Dipartimento di Scienze della Formazione, dei Beni Culturali e del Turismo, Universit\`a di Macerata, P.le Luigi Bertelli 1, 62100 Macerata, Italy      {\tt\small hagen.lehmann@unimc.it}}%
}

\begin{document}
\SetWatermarkAngle{0}
\SetWatermarkColor{black}
\SetWatermarkLightness{0.5}
\SetWatermarkFontSize{10pt}
% \SetWatermarkScale
% \SetWatermarkHorCenter
\SetWatermarkVerCenter{20pt}
\SetWatermarkText{\parbox{30cm}{%
\centering CognitIve RobotiCs for intEraction (CIRCE) Workshop\\
\centering IEEE International Conference On Robot and Human Interactive Communication (RO-MAN) 2020}}

\maketitle
\thispagestyle{empty}
\pagestyle{empty}

\begin{abstract}
This paper presents the first study in a series of proxemics experiments concerned with the role of personal spatial zones in human-robot interaction. In the study 40 participants approached a NAO robot positioned approximately at participants’ eye level and entered different social zones around the robot (personal and intimate space). When the robot perceived the approaching person entering its personal space, it started gazing at the participant, and upon the intrusion of its intimate space it leaned back. Our research questions were: (1) given the small size of the robot (58 cm tall), will people expect its social zones to shrink by its size? (2) Will the robot behaviors be interpreted as appropriate social behaviors? We found that the average approach distance of the participants was 48 cm, which represents the inner limit of the human-size personal zone (45-120 cm), but is outside of the personal zone scaled to robot size (16-42 cm). This suggests that most participants did not (fully) scale down the extent of these zones to the robot size. We also found that the leaning back behavior of the robot was correctly interpreted by most participants as the robot’s reaction to the intrusion of its personal space; however, our implementation of the behavior was often perceived as “unfriendly”. We will discuss this and other limitations of the study in detail. Additionally we found positive correlations between participants' personality traits, Godspeed Questionnaire subscales, and the average approach distance. The technical contribution of this work is the real-time perception of 25 keypoints on the human body using a single compact RGB-D camera and the use of these points for accurate interpersonal distance estimation and as gazing targets for the robot.
\end{abstract}

\section{Introduction and Related Work}

In order to ensure high reliability and behavioral functionality during collaborative tasks between humans and robots, it is important to take the behaviors that shape human-human interaction into consideration when designing behaviors for robots. An important part of human communication consists of nonverbal behaviors, specifically in cases in which intent and/or internal states are to be transmitted to the other.   
%when designing interaction behaviors for robots. 
%Despite recent scientific and technological advances in robotics, research in this field has not yet lead to widespread successful use of robots in operative contexts requiring close human-robot interaction. On one hand, this situation can be ascribed to technical difficulties. On the other hand, it can be related to a diffused expectation that robots need to be near perfect in their interaction behaviors in order to perform well when engaged in social exchanges with humans. This assumption shows its limits when we consider that humans are not perfect social agents either, and, to achieve common goals during human-human interactions, they mutually correct their respective behaviors to increase their possibilities of success. This solicits a change of perspective on the issue of how to build robots capable of good performance in effective human-robot interactions (hereafter HRI). A crucial part of this change of perspective is to understand what kind preconceptions and expectations humans have towards robots, based on the perception of themselves. 
These nonverbal cues are crucial in avoiding misunderstandings and ensuring a high efficiency while being engaged in collaborative tasks that require close physical interactions with other humans. Nonverbal expressions include conscious movements, such as gestures and touch, but also unconscious movements, such as body posturing, and are part of a research field called proxemics. Proxemics studies how humans use the space around them, specifically during social exchanges \cite{hall1968}. Humans are very sensitive to the intrusion of others into their different \textit{Personal Spatial Zones} (hereafter PSZ), making it possible to determine the average comfortable interaction distance for different levels of familiarity depending on the cultural backgrounds of the people involved \cite{lambert2004body} (e.g., PSZs for northern Europeans in Table~\ref{table:PSZ_Lamert}). Stratton et al.~\cite{Stratton1973personal} for example found in their study the mean approach distance between humans to be 0.51m.
%; this distance was increased to 0.56m when a tailor's dummy was used instead of another human. 
However the effectiveness and comfort of interactions in a dynamic environment depends not only on the distance of the other, but also on the approach speed and angle of the other \cite{mertens2014human}.

\begin{table}
\label{table:PSZ_Lamert}
\begin{center}
 \begin{tabular}{| c | c | c |} 
 \hline
 \textbf{PSZ} & \textbf{Range} & \textbf{Situation} \\ [0.6ex] 
 \hline
 Close Intimate & 0 to 0.15m & Lover or close friend touching \\ 
 \hline
 Intimate Zone & 0.15m to 0.45m & Lover or close friend only \\
 \hline
 Personal Zone & 0.45m to 1.2m & Conversation between friends \\
 \hline
 Social Zone & 1.2m to 3.6m & Conversation between non-friends \\
 \hline
 Public Zone & 3.6m + & Public speech \\ %[1ex] 
 \hline
\end{tabular}
\caption{Human Personal Spatial Zones (PSZ) for northern Europeans according to \cite{lambert2004body}.} 
\end{center}
\end{table}

%, specifically in the context of home companion robots \cite{walters2009empirical}. It has been shown that people react towards robots in similar ways they would react to humans and the many psychological factors influence the successful outcome of such an interaction (e.g.,\cite{mumm2011human,obaid2016stop,walters2005influence}).

Since robots that do not show appropriate distancing behavior may be perceived as threatening, or their "intent" may not be understood by their human counterparts \cite{mutlu2008robots}, the factors involved in human proxemics have increasingly become the subject of Human-Robot Interaction research (hereafter HRI). The majority of this research involves mobile robots on wheels and questions of social-aware navigation (e.g., \cite{huttenrauch2006investigating,rios2015proxemics,koay2017initial,mead2017autonomous}). It has been found that different factors modulate the distance people naturally assume from robots, or perceive as appropriate when a robot approaches them. The robot's physical characteristics (appearance: mechanoid vs. humanoid \cite{walters2009empirical}; height: effect not confirmed in \cite{Syrdal2007personalized}), physical characteristics in combination with behavior (PR2 robot with a reach of 0.92 m actively gesturing \cite{Mead2016perceptual}), and psychological aspects of robot behavior (mutual or averted gaze \cite{mumm2011human}; head direction \cite{takayama2009influences}). On the human part, it is the general attitude toward the robot (e.g., \cite{mumm2011human,obaid2016stop}), experience with robots \cite{takayama2009influences}, pet ownership \cite{takayama2009influences}. Finally, the overall context plays a part (robot approaching human \cite{walters2009empirical,Torta2011design} vs. human approaching robot \cite{mumm2011human,takayama2009influences,walters2009empirical}; direction of approach \cite{Torta2011design, walters2009empirical}; passing an object \cite{walters2009empirical}); Mead and Mataric~\cite{Mead2016perceptual} take a broader perspective and discuss the functional implications of interpersonal distance on speech and gesture production or perception. Most of the time, the interaction occurs while standing. Obaid et al.~\cite{obaid2016stop} explicitly study the effect of posture---robot or human standing vs. sitting. 
The mean approach distances found in different contexts vary. H{\"u}ttenrauch et al.~\cite{huttenrauch2006investigating} concluded that in HRI user trials most participants kept interpersonal distances from the robot corresponding to Hall's Personal Spatial Zone (0.45 m to 1.2 m). Walters et al.~\cite{walters2009empirical} report 0.57 m as the average, subject to a range of modifying factors; Takayama \& Pantofaru~\cite{takayama2009influences} report approach distances between 0.25 and 0.52 m. 

Interpersonal distances are emergent from the interaction of two or more actors and their individual personal spatial zones. One fundamental question that remains, to our knowledge, unanswered, is to what extent the human-robot distances observed are to be  attributed to participants' own PSZ and to what extent people imagine the robot having its own social zones that should be respected. Is it determined by who is approaching or who is being approached? It seems likely that people would often mostly care about their comfort rather than the robot's. Therefore, in this work, we have explicitly prepared a scenario in which the robot wants to signal that the interlocutor has entered his personal/intimate zone. We designed two types of behaviors---gazing and leaning back---to signal intrusion into the personal and intimate zone, respectively. A natural response to someone entering our intimate zone is stepping or leaning back to increase the distance again. After experimenting with the stepping back behavior on the Nao robot, we found it quite slow and unnatural. Therefore, for this work we implemented the leaning back behavior which is, to our knowledge, quite new in an HRI context (cf. \cite{Shiomi2018} though). 

%In our research we wanted to test whether the size of the robot has an influence on the size of the personal spatial zones that people expect the robot to have around it. In other words, would people stand closer to smaller robots after approaching them to interact with them? It has been shown that the size of an approaching robot has an influence on how comfortable people feel to interact with them \cite{hiroi2011influence}; 
%Second, most of the works reviewed above feature rather tall robots (over 1 m), like the PR2. 
Our general research questions were:
\begin{itemize}
    \item Do people expect a robot to care about its personal spatial zones?
    \item Do people expect robot personal spatial zones to scale to the robot size?
    \item Which robot behaviors will be correctly interpreted as the robot signaling discomfort about its personal spatial zones being invaded?
\end{itemize}

To explore the answer to these questions, we used the humanoid robot Nao that is 58 cm tall. Torta et al.~\cite{Torta2011design} used the Nao but studied the opposite situation: robot approaching human. Obaid et al.~\cite{obaid2016stop}, also using Nao, investigated both situations: robot approaching human and human approaching robot and the effect of posture---standing vs. sitting---on the interaction. For the conditions where the human was the active participant, the distance left when the robot was sitting was on average 0.35 m whereas the distance left when the robot was standing was on average 0.5 m. The distance was similar when the robot approached the human, with 0.1 m larger distance that a standing human tolerated (as opposed to sitting human). However, in all these cases, the Nao robot was on the ground and thus very small compared to the human which may have biased the results. Therefore, in our experiments we have elevated the Nao to be comparably tall to the participants. We have implemented two versions of the PSZs---human scale and scaled down to robot size---and triggered the signaling behaviors based on these and evaluated how this was perceived by the participants.

In the remainder of the paper, we will discuss how we implemented these behaviors and how the experiment was designed. After this, we will present our results and discuss them from the perspective of our research questions.

\section{Methods}
\label{section:exp_setup}

%HL Changed here
The experiment took place at the Department of Cybernetics at CTU in Prague, all participants were native Czech speakers. Therefore, the forms, instructions and questionnaires we presented to the participants were in Czech. The written and verbal information given by the participants were transcribed and translated into English for analysis by a bilingual native Czech speaker.
Before the experiment started, the participants gave signed consent, which included instructions about the experiment. Additionally, we asked them for their age and their experience with robots (on a 5-point Likert-Scale), and gave to each participant the Ten Item Personality Inventory (hereafter TIPI) \cite{gosling2003}. 
Our sample consisted of 40 naive participants in a within-subject design, in which each participant interacted with the robot in 3 consecutive sessions. In each session, a different condition was presented to the participant. The order of conditions was randomized. The participants were not given any information about the purpose of the study or the reasons behind the robot’s behavior. A version of the Godspeed questionnaire was given to them after every session. One additional custom-made questionnaire was completed after the end of the last session.

\begin{figure}
    \centering
    \includegraphics[width=0.45\textwidth]{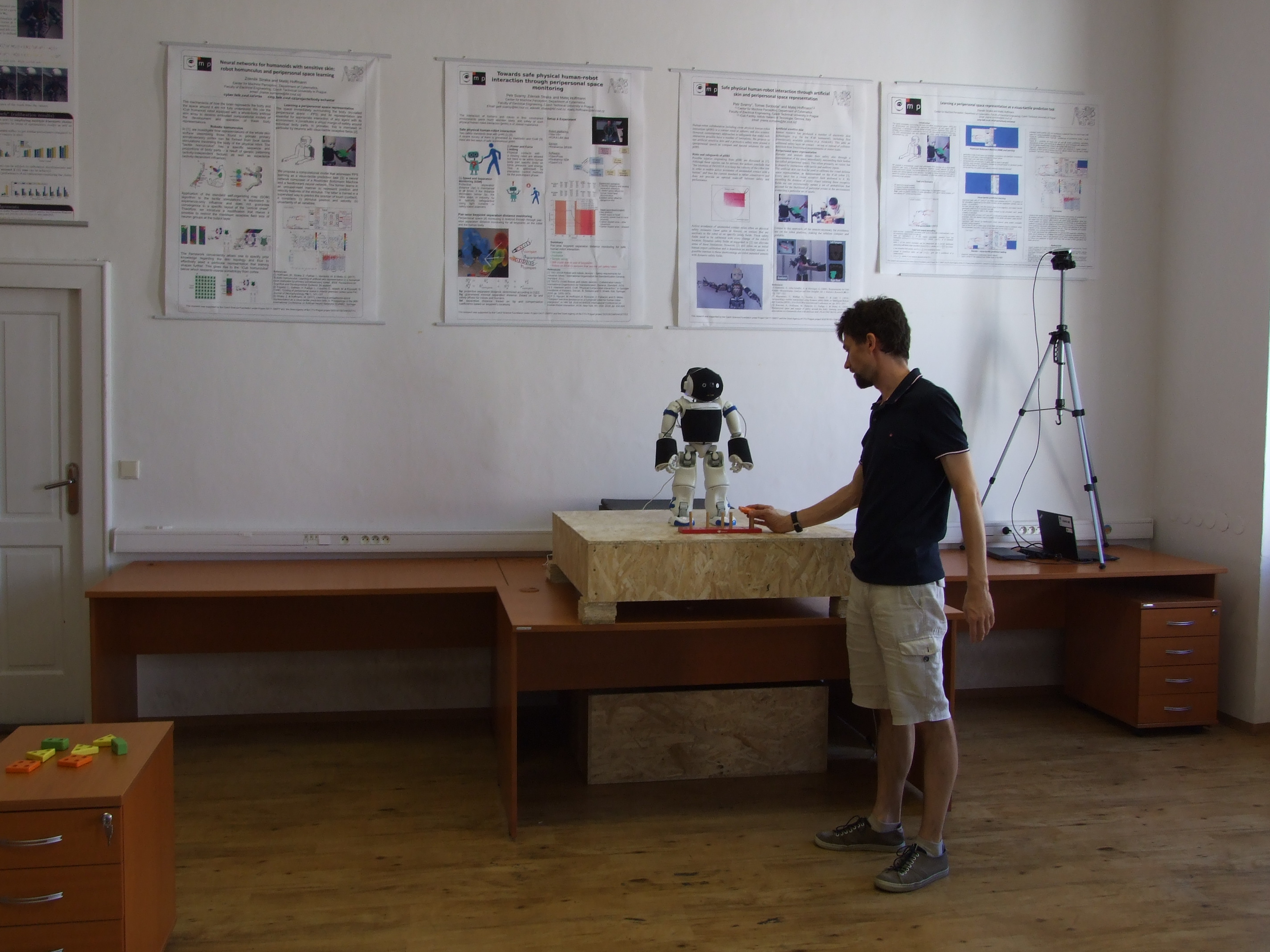}
    \caption{Experimental setup.} 
    \label{fig:setup}
\end{figure}

The robot was positioned on a platform (Fig.~\ref{fig:setup}) such that its height could be adapted to match that of the participants---target height of the robot from the ground corresponding to 20 cm less than the participant's height. This was a compromise to comply with two conflicting objectives: (i) similar overall height of robot and participant, (ii) the possibility of leaning over to the robot over the platform's edge. The height adjustment became necessary due to the small size of the Nao robot: to avoid that participants have to crouch to interact with the robot, it was placed on a table.

%\footnote{For participants under 160 cm, the platform was removed completely and the robot (59cm) was placed directly on the table (75cm). For participants around 165cm tall, the platform was used, meaning that the robot total height was 155cm, deviating somewhat from the goal.}

The overall setup can be seen in Fig.~\ref{fig:setup}. Most of the features detailed in this section are also illustrated in this video \url{https://youtu.be/gvICAkfK2CA} and the code employed is available at this public repository~\cite{github-nao-hri-personal-zones}.

A typical session had the following structure. The participant was asked to enter the experiment room, cross it, go to the robot and to read the instructions written on the robot’s chest and to follow them. For each condition, there was a different set of instructions. The activity described in the instructions consisted of a color and shape matching game, in which blocks with three different shapes and colors had to be put in the correct position on a stand in front of the robot. In pilot experiments, we found that the game was sufficiently complex to keep the attention of the participants over the three sessions. 

\begin{figure}
  \begin{center}
    \includegraphics[width=0.45\textwidth]{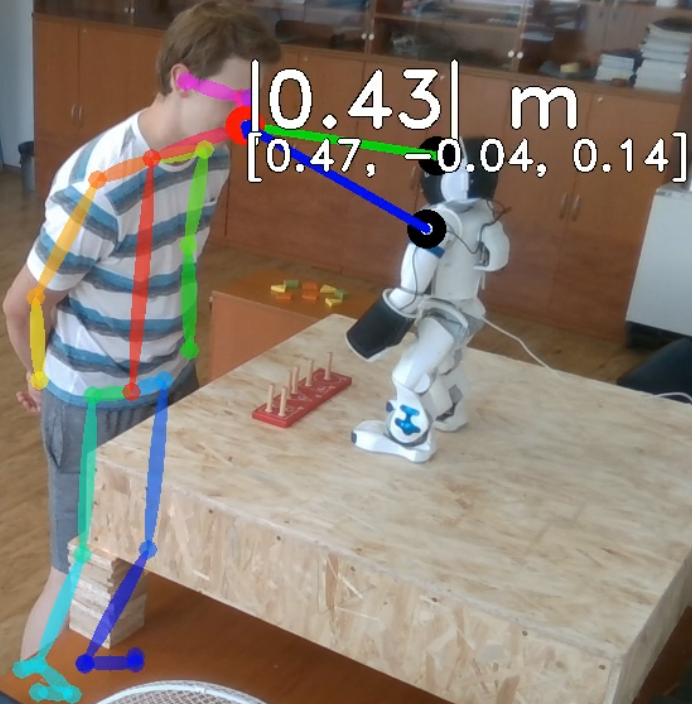}
    \end{center}
  \caption{Human keypoints detection for distance measurement and gazing. Green arrow -- head gaze target for the robot. Blue arrow -- closest human-robot keypoint pair (nose to chest), at a distance of 0.43m. The coordinates are those of the human nose in the robot base frame (robot waist).} 
  \label{fig:keypoints}
\end{figure}

\begin{figure}
\centering
  \includegraphics[width=0.45\textwidth]{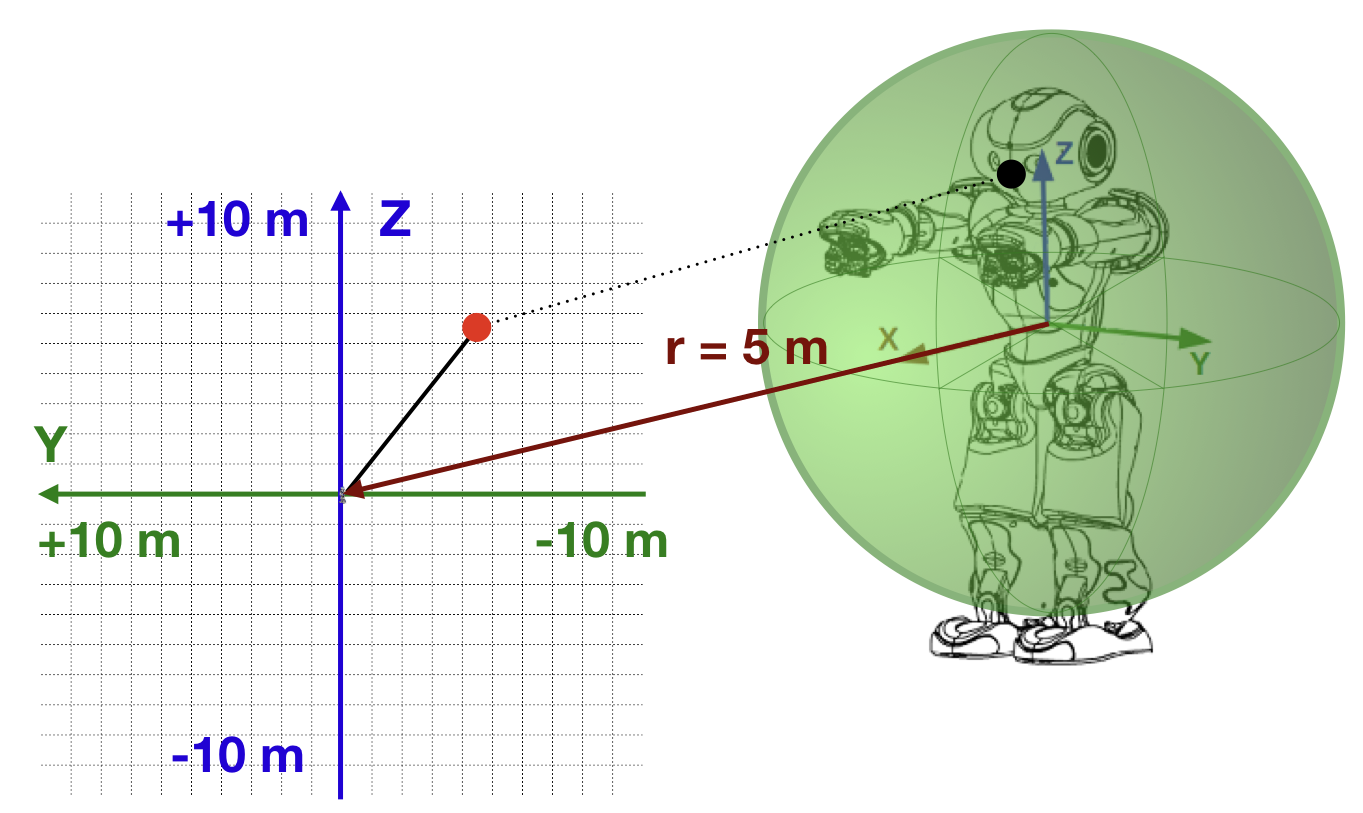}
  \caption{Targets for random head gaze of Nao. Every 5 seconds ($\mathcal{N}(5,1)$), a new target was chosen at random (uniform distribution) from a $20\times 20$ meters virtual plane placed $5$ m in front of the robot. Robot picture from http://doc.aldebaran.com.} 
  \label{fig:nao_random_gaze}
\end{figure}

\textbf{Robot behaviors.} The behaviors the robot displayed in the three different conditions were the same—gazing and leaning back—but the context in which they were triggered was different. Apart from the control condition in which they were started randomly, the robot started to gaze at the participant when she entered its personal zone in order to acknowledge that it has seen the person \cite{harrigan2005proxemics,fiore2013toward}, and to lean back when the participant entered its intimate zone in order to indicate that the person is getting too close. The leaning back behavior is based on findings from Human-Robot Interaction research about pre-touch reactions \cite{Shiomi2018}, simulating the reaction of a human that would experience such an intrusion from a stranger.
\begin{itemize}
    \item Head gaze: For the head gaze we used the two degrees of freedom in the robot neck (yaw and pitch) to look at a point in 3D space in front of the robot. In the Control condition, the target was on virtual plane placed in front of the robot and a new one was picked every 5s -- see Fig.~\ref{fig:nao_random_gaze} for details. In the \textit{Robot} and \textit{Human} conditions, the head was commanded to gaze at one of the keypoints on the participant’s head—provided she was in the corresponding personal zone of the robot. If available, the nose keypoint was chosen (green arrow in Fig. ~\ref{fig:keypoints}); if nose was not detected, another keypoint on the head was chosen.
    \item Leaning back: The leaning back behavior was designed by hand using Choregraphe’s timeline editor. The robot leans around $18$ degrees back in the hips from its standing position and the whole action takes $1.04$ seconds; arms move in opposite direction to maintain balance.
\end{itemize}

% and then the TIPI -- if we decide to actually use it for anything
\textbf{Hypotheses.} In order to answer our research questions and to explore whether humans expect personal spatial zones around robots to be similar to their own spatial zones, or whether they expect the size of these personal spatial zones to correspond to the size of the robot, our experimental design was informed by the following hypotheses:

\begin{itemize}
    \item H$_{0}$: Humans expect the personal spatial zones around a robot to correspond to the robots size
    \item H$_{1}$: Humans expect the personal spatial zones around a robot to correspond to the size of human personal spatial zones.
    \item H$_{2}$: Humans are able to recognize proxemics behaviors exhibited by robots to be associated with the intrusion of the robots personal spatial zones.
\end{itemize}

\begin{figure}
\centering
  \includegraphics[width=0.45\textwidth]{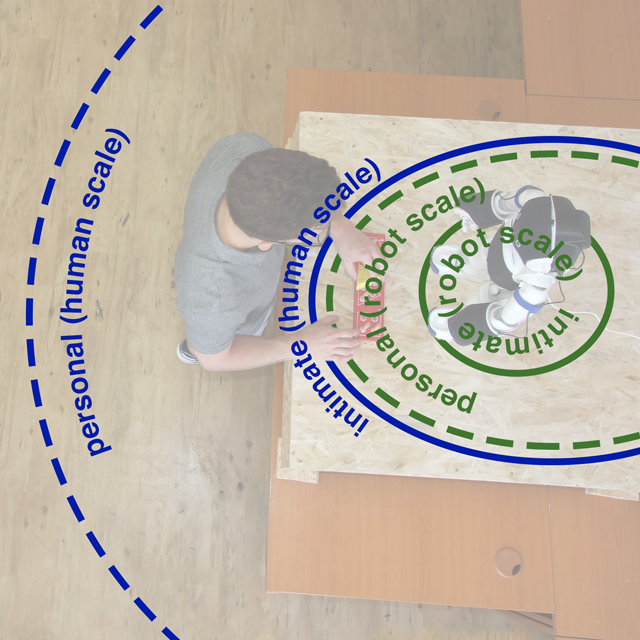}
  \caption{Experimental setup -- top view. Personal spatial zones: human scale (blue) and scaled by robot size (green).} 
  \label{fig:setup_with_zones}
\end{figure}

\textbf{Conditions.} In each condition the robot exhibited the above described proxemics behaviors. The key difference between the conditions was the distance between the human and the robot, at which the robot started to exhibit these behaviors. The different distances corresponded to either human personal spatial zones, or were scaled down to the size of the robot -- Fig.~\ref{fig:setup_with_zones}.

\begin{itemize}
    \item Condition 1 (Control): Upon the approach of the participant, the robot exhibited the behaviors head gaze movements and leaning back movements randomly. The leaning back was triggered every $20$ seconds for $5$ seconds. For the gaze, it was every $5$ seconds with a standard deviation of $1$ second. These settings were chosen so that the quantity of the robot behaviors was roughly similar to the “human scale” condition. The period for the leaning back behavior was chosen such that the behavior is triggered around 2-3 times during the experiment, drawing on pilot experiments. 
    \item Condition 2 (Robot scale): The behaviors of the robot (head gaze movements, leaning back) were triggered when the human entered the corresponding spatial zones around the robot scaled to the robot's size: 16 cm for the intimate zone; 42 cm personal zone. 
    \item Condition 3 (Human scale): Same as Condition 2 with zone boundaries according to human size (45 cm intimate, 120 cm personal -- see Table \ref{table:PSZ_Lamert}).
\end{itemize}

\begin{figure*}
  \begin{center}
    \includegraphics[width=0.9\textwidth]{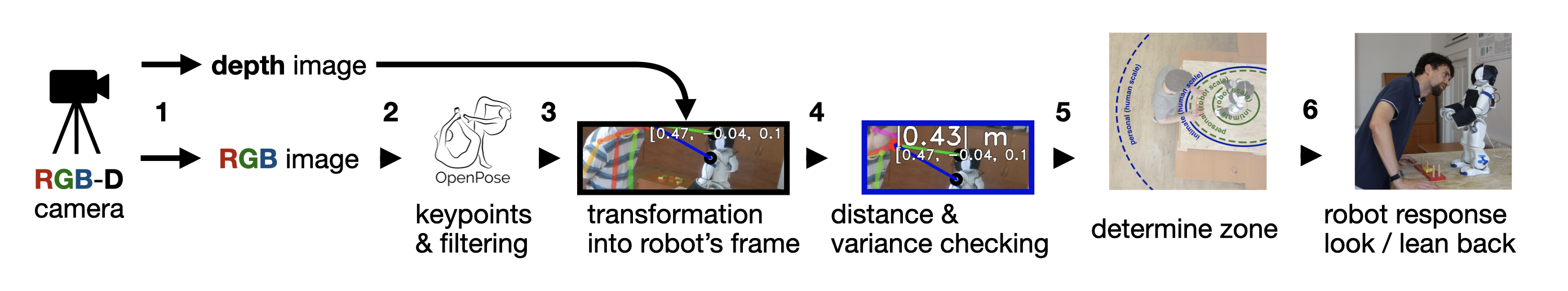}
    \end{center}
  \caption{Software architecture. Participants interacting with the robot are perceived using an external RGB-D camera (1). The RGB image is fed to OpenPose~\cite{Cao2018} which estimates human keypoints in the image, applying confidence thresholds (2). The image is then fused with the depth information to acquire 3D coordinates of the human keypoints, which are then transformed to the robot frame of reference (3). The distances are checked for consistency (4) and used to determine the personal spatial zone in which the participant is located (5). Appropriate robot behaviors are triggered (6). The complete software is available at \cite{github-nao-hri-personal-zones}.
  % For the filtering - first is just openpose threshold; second, the distances variance is checked (<.1, experimentally deduced value) + if more than 2 keypoints were found
  } 
  \label{fig:software}
\end{figure*}

\textbf{Real-time interpersonal distance measurement.}
In order for the robot to exhibit the correct behaviors depending on the condition that was tested, reliable online measurements of the separation distance between the participant and the robot were required. The position of any part of the robot is readily available from joint encoders and forward kinematics. For tracking participants' poses, we employed the following pipeline: color images from a camera were fed via the Python API to the OpenPose library \cite{Cao2018} to calculate the estimated human keypoints' position in the image -- see Fig.~\ref{fig:keypoints}. As positions in 3D were needed, we used an RGB-D camera (Intel® RealSense™ D435) rather than the Nao head camera. The resulting keypoint locations were then deprojected using the aligned depth image, thus receiving 3D coordinates in the camera's frame of reference. This camera was external and the coordinates were finally transformed into the robot's frame of reference. The pipeline is schematically illustrated in Fig.~\ref{fig:software}. Note that the camera is small and lightweight and could be easily attached on top of the Nao's head, for example. However, we did not choose this solution in order not to influence the robot's appearance (anthropomorphism). 

%Following we needed a coordinate transformation from the camera to robots body, that would be reliable and fast to set up as robots position varies. We used multiple snapshots of the robots hand position with detachable Aruco marker on it combined with the marker centre position in 3D from the camera, which gave us enough data to extract the transformation into NAOs coordinate system.

%For person detection, we had used OpenPose on the 2D image, which gave us keypoints for the entire body.

%External GPU was needed to run OpenPose real-time, so we used Lenovo Thunderbolt™ 3 Graphics Dock with NVIDIA\textsuperscript{\textregistered} GeForce\textsuperscript{\textregistered} GTX 1050 (4 GB GDDR5) as it was already available to us. That got us to roughly $15$ FPS, which was feasible.

%\subsubsection{Distance Measurement}
%PPS distance is usually measured for the whole body, so we decided to use all head keypoints and mid hip keypoint only. From that, we could calculate the distance between the participant and the robot. When the variance between those distances exceeded $0.1$ m$^{2}$, or we got less than three keypoints, we ignored the frame.

%We used minimal distance from the robot to the participant keypoints to determine the currently intruded zone of the robot. The robot had two keypoints, one on its startup button, which acts as a torso and the other on its top camera.

The keypoints thus obtained were used to calculate the interpersonal distance between the participant and the robot. As the PSZs have a rather holistic or whole-body character, we decided not to use all the possible keypoint pairs, ignoring the limbs in particular.\footnote{Walters et al.~ \cite{walters2009empirical} considered trunk parts only, ignoring arms and regarded this as comparable to Hall or Stratton et al.~\cite{Stratton1973personal}.} Instead, we used two keypoints on the robot (chest and head). For the human, we considered the keypoint in the waist area and then the head keypoints. These robot and human keypoints formed pairs, for which all the distances were computed and the closest taken as the current interpersonal distance (Fig.~\ref{fig:keypoints}). Most of the time, the shortest distance was between the robot's head and one of the keypoints on the human's head (typically nose).

\textbf{Other Measurements.}
\label{section:measures}
%HL - changed here
The distances between participants and the robot---more specifically, the distances between pairs of keypoints on the human and the robot---were recorded throughout every experiment. The logs and videos with keypoints and distances overlaid were used for subsequent analyses.
With the goal of assessing the personality dimensions of our participants we asked them at the beginning of the experiment to complete the Czech version of the TIPI \cite{sefara2015socio}.

In order to evaluate the impression the participant had of the robot's behavior we asked them after each condition to complete subscales I, II and III (Anthropomorphism, Animacy, and Likeability) of the Godspeed Questionnaire \cite{bartneck2009}. We chose these three subscales because they seemed most relevant for answering our research questions. 
%The internal consistency of the individual subscales of the Godspeed questionnaire and their separate usability has been shown in various previous studies (e.g. \cite{Lehmann2015}).   

Additionally, at the end of the experiment, we used a short, custom-made questionnaire in order to evaluate the participants' general experience during the experiment. In this questionnaire we asked the participants whether they noticed a difference between the conditions, and if yes, what, according to them, this difference was. We also asked in which condition they found the gaze behavior and the leaning back behavior of the robot most appropriate. For the leaning back behavior, we additionally inquired how the participants interpreted it. The last item of this questionnaire was a request for general feedback and comments on the experiment. After the participants finished the experiment (including the questionnaires) we did a short structured interview, in which we asked the following questions ``What did you like and what didn't you like about the experiment?'', ``What do you think about the leaning back behavior of the robot?'', and ``How would you describe the behavior of the robot in general?''. 

%\subsection{Ethics Statement}
%This research project has been officially approved by the Czech Ministry of Education and by the Faculty of Electrical Engineering of \v{C}VUT. 

\section{Results}

\subsection{Characteristics of the Sample} \label{Participants}
Our sample consisted of 40 participants (22 female, 18 male; mean age 31.5 years, ranging from 18 to 62). On a 5-point Likert scale concerning their experience with robots and ranging from 1 = no experience to 5 = very experienced, they reported an average experience with robots of 1.6. The subjects were thus largely naive with respect to experience with robots.

\subsection{Distance from the robot}
We recorded the distance the participants kept from the robot. For evaluation, the minimum distance is often used \cite{bailenson2001equilibrium,mumm2011human}. However, this is not appropriate in our situation, as the distance is also determined by the game and the participants are asked to explicitly lean toward the robots at some point. We were interested in the ``natural'' distance the participants assumed for the interaction with the robot. In our context, this was the distance at which they stopped after approaching the robot in order to read the instructions on the robot's chest. This information was extracted from the video recordings with the keypoint distances overlaid (see Fig. \ref{fig:keypoints}). On average our participants stayed away from the robot 47.7cm with a standard deviation of 11.3 cm. We found no significant differences between the distances in the different conditions ((F(2, 114) = 0.48, p = 0.62, $\omega^{2}$= -0.01)), albeit there was a small drop from the first trial to the last (average 49.3 over 47.3 to 46.6 cm) which may be due to familiarization with the robot (data lumped across robot behavior condition, the order of which was randomized).\footnote{Walters et al.~\cite{walters2009empirical} report a much greater adjustment factor of 13 cm on first encounter. } 

For individual participants, the standard deviation of the distances they assumed in the three sessions was 4.9 cm, i.e. smaller than the st. dev. across participants, indicating that every individual had her preferred separation distance from the robot.  
Females kept an average distance of 48.9 cm and males of 45.6 cm, but this difference did not reach significance (t(37)=1.05, p=0.3).
%For both groups the st. dev. was 10cm.  

\begin{figure}
  \begin{center}
    \includegraphics[width=0.45\textwidth]{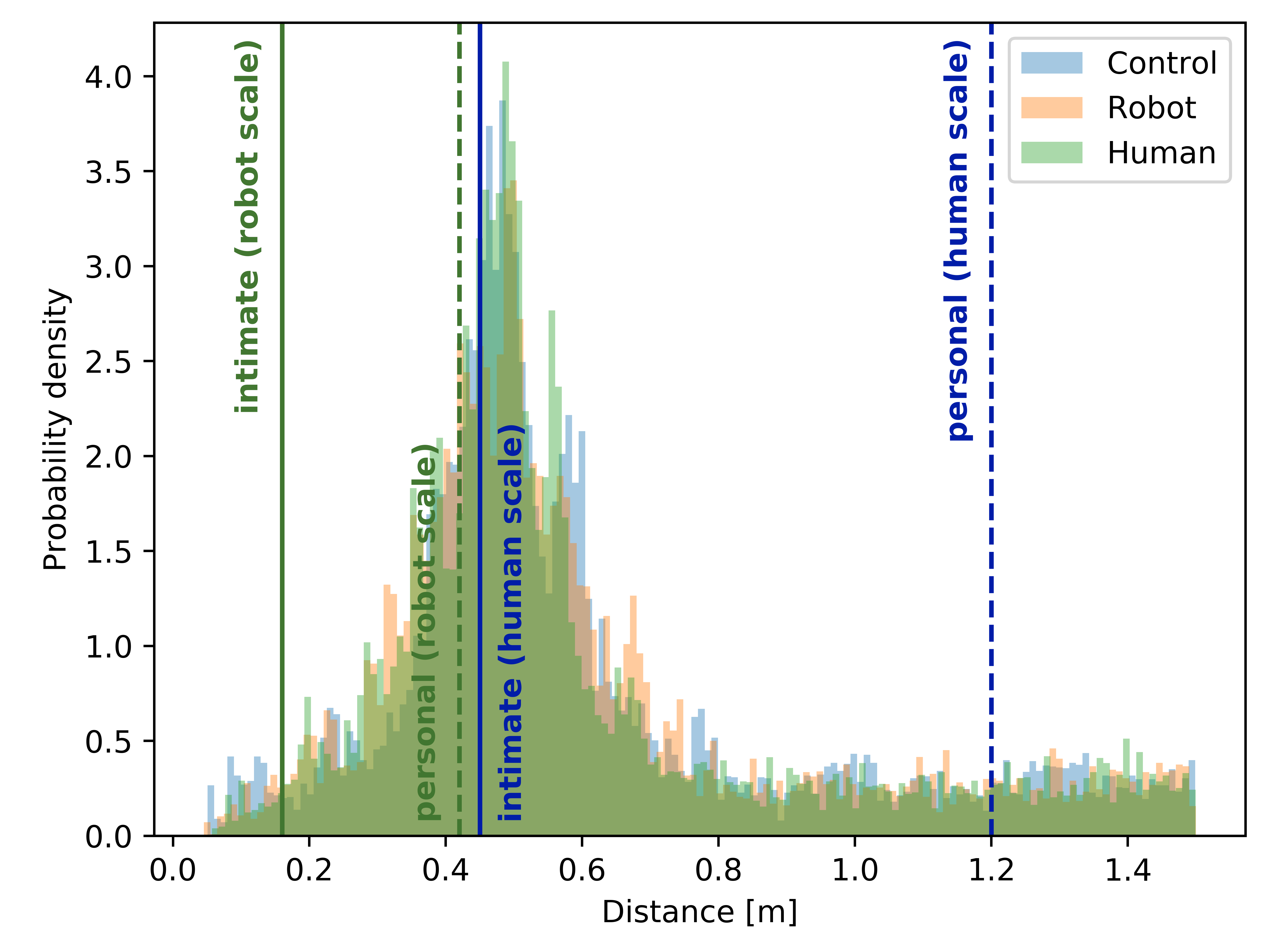}
  \end{center}
  \caption{Statistics of distances of participants from the robot. Aggregate statistics from all experimental runs; different Conditions (Control/Robot/Human) color coded. Distances extracted from logs of keypoint-to-keypoint distances. Distances over 1.5m truncated.} 
  \label{fig:distances}
\end{figure}

Additionally, we used the logs of keypoint-to-keypoint distances to investigate the distances assumed during entire experiments. The distribution is shown in Fig.~\ref{fig:distances}. This result may be affected by the experimental condition, as the robot behavior is different and dependent on the distance from the participant. However, the plot is not suggestive of any strong effect---albeit a small tendency for lower density in the human condition at around 47 cm is apparent (robot's leaning back may have induced a larger distance assumed by the participants). All distributions peak at around 50 cm from the robot, indicating that most participants would not enter the standard intimate zone of the robot interlocutor (human scale -- 45 cm). 

\subsection{Results for Godspeed Questionnaire Subscales}
The descriptive statistics for the three Godspeed Questionnaire subscales can be found in Table~ \ref{DescriptiveStats}. For the Anthropomorphism dimension, participants overall scored the robot higher than a ``neutral'' score of 3 only in condition Robot scale, while the participants scored the robot lower than this ``neutral'' score in both conditions Control and Human scale. 

\begin{table}
\begin{center}
\includegraphics[width=0.45\textwidth]{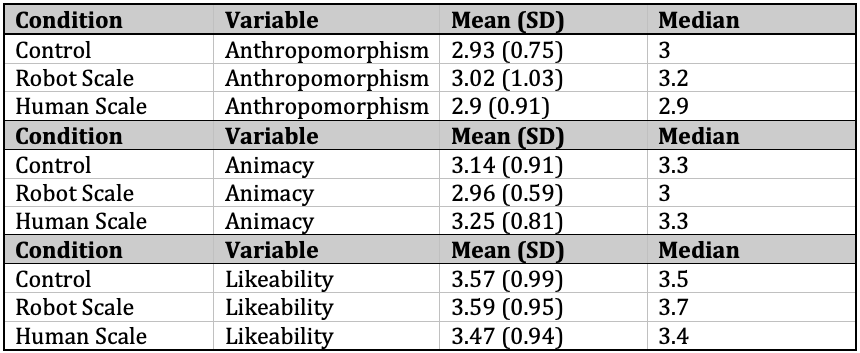}
\end{center}
\caption{Descriptive statistics for Anthropomorphism, Animacy and Likeability.} \label{DescriptiveStats}
\end{table}
%Participant 019123 was excluded - during first trial, participant first assumes a huge distance of 1.2m – probably waiting to be talked to by the robot. Experimenter then reminds her to go and read the instructions – then she moves to a distance of around 0.55m.

The effect of condition on participant ratings along this dimension was assessed using a repeated measures ANOVA, finding no significant effect (F(2, 117) = 0.19, p = 0.83, $\omega^{2}$= -0.01).

For the Animacy dimension, participants would score the robot higher than a ``neutral'' score of 3 in the conditions Control and Human scale, and lower than this in condition Robot scale. The effect of condition on participant ratings along this dimension was assessed using a repeated measures ANOVA, showing no significant effect (F(2, 117) = 1.45, p = 0.24, $\omega^{2}$= 0.01).

For the Likeability dimension participants would score the robot higher than a ``neutral'' score of 3 in all three conditions. The effect of condition on participant ratings along this dimension was assessed using a repeated measures ANOVA, showing again no significant effect (F(2, 117) = 0.16, p = 0.85, $\omega^{2}$ = -0.01).

\subsection{Correlations between the measured variables}
In order to further explore the structure of the collected data we examined the correlations between the different measured variables. We calculated Pearson Correlation Coefficients to investigate potential correlations between the TIPI scores of our participants and the average distance they kept from the robot. We found a positive correlation between the agreeableness score and the distance (r(38) = .348, p = .028), and a weak positive correlation between the conscientiousness score and the distance (r(38) = .327, p = .04). We found no correlations between the experience with robots score and distance, and age and distance. We tested also for possible correlations between the distance and the Godspeed questionnaire answers of the participants. Here we found a positive correlation between the average (over all conditions) anthropomorphism score and the distance (r(38) = .5504, p = 0.0002, and a positive correlation between the average (over all conditions) animacy score and the distance (r(38) = .4337, p = .005). Even though this was not directly related to our research questions, we further tested for correlations between the experience with robots score and the Godspeed questionnaire results, and the Age of the participants and the Godspeed questionnaire results. For the experience with robots score, we found a negative correlation with the Likeability subscale (r(38) = -.322, p = .044); for age, we found a positive correlation with the Likeability subscale (r(38) = .369, p = .019).   

\subsection{Custom-made questionnaire and structured interview}
In the custom-made questionnaire, only one person out of 40 reported that she did not notice a difference between the conditions. 
For the gaze behavior, 10 participants rated it most appropriate in the Control condition, 13 in Robot scale condition, and 17 in the Human scale condition. For the leaning back, the counts were 8, 15, and 17 respectively. The participants thus rated both the gaze and the leaning back behavior most appropriate in the human scale condition and least appropriate in the control condition. %In the robot scale condition the appropriateness of the behaviors was rated in between. 
%The results for the ratings of the appropriateness of the gaze and the leaning back behavior show that in the control condition 10 participants rated the gaze behavior and 8 participants the leaning back behavior most appropriate. In the robot scale condition 13 participants found the gaze behavior and 15 the leaning back behavior most appropriate. In the human scale condition 17 people found both the gaze and the leaning back behavior most appropriate.

%\begin{table}
%\begin{center}
%\includegraphics[scale=0.37]{GraphsFigures/RatingAppropBehav.png}
%\end{center}
%\caption{Participant interpretation of the leaning back behavior of the %robot. \mh{THIS TABLE IS TAKING TOO MUCH SPACE. THE INFORMATION CAN BE %GIVEN IN TEXT PERHAPS?}} \label{RatingAppropBehav}
%\end{table}

Additionally we studied how the participants interpreted the leaning behavior. They were able to freely answer this question in the questionnaire. The majority of the answers fell into three categories: most participants (n=13) interpreted the behavior as a reaction to the intrusion of the intimate/personal space of the robot, followed by ``the robot was shocked'' (n=7), and ``the robot was afraid'' (n=5).
%One person interpreted the reaction as astonishment, one as acknowledgement, and another as displeasure. The rest of the participants didn't give an answer. 3 answered they didn't know. 

%\begin{table}
%\begin{center}
%\includegraphics[scale=0.37]{GraphsFigures/MeaningOfLeaning.png}
%\end{center}
%\caption{Participant ratings of in which situation was gaze behavior or %leaning behavior most appropriate} \label{MeaningOfLeaning}
%\end{table}

The last part of the custom-made questionnaire allowed the participants to post miscellaneous comments. In combination with the concluding structured interview, the main qualitative findings pertain to the leaning back behavior, which was overall perceived as unfriendly or detached. This had in part to do with a contradiction in the scenario, which was reported explicitly by some participants. As part of the interaction, they were asked to lean towards the robot and whisper to it, but the robot in this occasion would lean back. Additionally, some participants reported that the leaning back was too fast and that they got scared the first time it was triggered. Some participants also reported that they perceived the gazing behavior (human or robot condition) as being ``stared at'' due to the lack of facial expressions on the robot.

\section{Conclusion, Discussion, Future Work}
%\subsection{Discussion}
In summary, the average approach distance people assumed from the robot was 48 cm, which is very close to the distances between humans reported by Stratton et al.~\cite{Stratton1973personal} (51 cm) and at the inner boundary of human-size personal zone (45-120 cm), but outside of the personal zone scaled to robot size (16-42 cm), suggesting that most participants did not automatically scale down the extent of these zones to the robot size.
%In summary, we could show that when interacting with the robot, people kept an average distance that was slightly beyond the border of intimate space of human-human interaction, in the personal space. This could be an indication that people assume that robots have PSZs around them and that these zones are independent from the size of the robotic embodiment. Our findings further illustrate the importance of the context and speed for the use of leaning back as a potential pre-touch reaction in HRI. 
%Despite the limitations of this study, which occurred partly due to the exploratory nature of the research, 
%The results hold promise for further investigations into PSZs in HRI, where we plan to include also tactile interaction, as touch crosses the ultimate social zone: the skin.
%The results from the distance data showed that people stayed on average 48 away from the robot, which falls into the human scaled personal space zone.
This distance was independent from the gender of the participant and from the behavior of the robot. That is neither the gazing nor the leaning back affected the approach distance significantly. Mumm \& Mutlu~\cite{mumm2011human} could show significant effects of mutual gaze vs. averted gaze on the part of the robot; in our case, the gaze was either random (that is not explicitly averted) or mutual. The leaning back was perceived as unfriendly by some participants, but did not translate into important changes in interpersonal distance.

It should be noted that the measures reported are constrained by two additional
factors. First, the edge of the table and platform on which the robot was placed was around 40cm horizontally from the robot keypoints (head, torso)---creating a barrier for the participants, unless they leaned over to the robot (which many did). From the results of the distances analysis, this did not seem to importantly affect the experiment. Second, the distance was also affected by the eyesight of the participants’---a few participants had to go close to the robot in order to read the instructions.
% Adjustment factors: \cite{walters2009empirical} -7 giving object; reading instructions

The analysis of the results from the Godspeed questionnaire showed no differences between the conditions---contrasting somewhat with the results obtained from the custom-made questionnaire and the information collected during the structured interview after the experiment, in which all but one participant reported to have noticed differences and described the differences in detail. We hypothesize different reasons for this lack of congruence. According to the participants' statements, our implementation of the leaning back behavior was problematic. Although it was correctly interpreted by most of them as being a protective reaction of the robot towards an intrusion of its personal (intimate) zone, the behavior's specific kinematics seem to not have been appropriate. Firstly, participants found it too fast and perceived it as too abrupt. This made the robot appear shocked or afraid of the participant. Second, it conflicted to some degree with the task the participants were asked to do: approaching the robot and whispering to it, during which the robot reacted by leaning back abruptly. This has startled some participants and was perceived as unfriendly. The third issue with the leaning behavior was more technical: some participants stood in front of the robot exactly at the border between the personal and the intimate zone, resulting  in the robot oscillating between leaning back and forth. This issue seems relevant for any situation in which for example safety zones have to be implemented to ensure a secure interaction between robots and humans. This problem can be mitigated by some filtering approaches (low-passing) to the distance processing around the zone boundaries to slow down the dynamics. However, delays on the robot behavior when, say, the intimate zone is intruded, may not be perceived positively either. An asymmetrical solution---withholding the reaction when the distance increases---may be a possibility. 
%Potential solutions to this problem can be the use of continuous instead of discrete behaviors, and the use of tactile sensing. 

The use of the TIPI allowed us to analyse the data for potential correlations between the psychological variables like self-assessed personality dimensions and the distance people kept from the robot. We found positive correlations between the agreeableness and conscientiousness personality traits of our participants and the distance they kept from the robot. A high score in agreeableness is linked to being considerate and kind; a high score in conscientiousness is linked to being careful and having the desire to do a given task well. It could be argued that people with these personality traits tend to be more cautious when entering a new potential social interaction, more specifically when intruding someone's personal space.
%The more agreeable people rate themselves, the further they stay away from the robot.
%The more conscientious people rate themselves, the further they stay away from the robot.
We also found positive correlations between the average scores of two Godspeed Questionnaire subscales, Anthropomorphism and Animacy, and the average distance people kept from the robot. Both of these subscales deal with how our participants perceived the robot. It could therefore be argued that the more people perceived the robot as human-like in its appearance and movements, the more the participants behaved towards it---from a spatial perspective---like they would towards a human.
Further, we found a negative correlation between the experience of our participants with robots and the average score of the Likeability subscale of the Godspeed questionnaire. This signifies that, in our experiment, the people that were less familiar with robots found the robot more likeable. Another positive correlation between the age of our participants and average score of the Likeability subscale points in the same direction: The older people were, the more likeable they found the robot. Since our participant sample was quite diverse ranging from young students of computer science to people over 60 from outside academia, these two results can be seen as coherent. It seems that---at least in our experiment---experienced people have a more realistic and technical perspective on the capabilities of the robot.
%\mh{, expecting less from it and finding it less likeable.}. \mh{I ADDED THESE COUPLE OF WORDS THAT FOR ME CLOSE THE ARGUMENT SOMEWHAT. WOULD THAT DO?} 
%\hl{This works great.}
%\mh{where does this come from?}  
%\hl{hm. I think this is a possible interpretation of the correlation between likeability and age. The older people are somewhat more naive about the robot and imagine that it can do much more than it actually can. Whereas the CS students are more realistic and see it as an mechanical devise - less likeable. But we can leave this out, or give another explanation.}  

One limitation of the present study is that the robot size could not be manipulated. We plan to continue this research adding the Pepper robot to further investigate the potential effect of robot size on the human perception of robot PSZs. Using Pepper will allow us to repeat our study with a robot that has a size roughly halfway between the NAO robot and a human adult. It will further make it possible to explore the effect of the speed of the leaning back behavior, since Pepper can smoothly move backwards when a participant leans in. Additionally, Pepper will allow us to simulate to some extent changes of facial expression via eye color. It was pointed out by some participants that the static face of the NAO we used was perceived as staring.
As pointed out above, the biggest limitation of this study was most likely the design of the leaning back behavior. In the future, we will adjust the speed of the behavior and make it more continuous. %Where people stopped in front of the robot might have been influenced by two additional factors. First, by the edge of the platform on which the robot was placed in order to be aligned with the participants. However, judging from the distances results, this did not seem to importantly affected the experiment. Second, the distance might have also been affected by the eyesight of the participants.
%---few participants may have had to go close to the robot in order to read the instructions.
The knowledge gained in this and the planned follow-up studies can flow into behavioral modules for social robots; such a module monitoring the personal space and leading onto ``anxiety'' of the robot if its personal/intimate space was invaded, has been very recently developed in \cite{maniscalco2020ass4hr}.

Finally, a contribution of this work is also that we demonstrated that a single compact lightweight RGB-D camera and a laptop is sufficient for reliable real-time perception of the position of the human with respect to the robot. Moreover, thanks to the human keypoint extraction pipeline, 3D positions of 25 points on the human body (joints on the body, plus nose, eyes, ears) are available (our code is available at \cite{github-nao-hri-personal-zones}). Although such a resolution is normally not regarded as necessary in proxemics research \cite{Stratton1973personal,walters2009empirical}, it opens up new possibilities. In our scenario, availability of 3D positions of keypoints on the human head allowed us to get a more accurate estimation of interpersonal distance. At the same time, they were simultaneously used as targets for the robot's gaze. Availability of detailed information---3D positions of human keypoints from the camera and robot keypoints from forward kinematics---can be exploited too. This has been demonstrated on the iCub humanoid robot \cite{Nguyen_HRI_2018}---employing stereo cameras in the eyes instead of external RGB-D camera---but in the context of robot monitoring its and the interlocutor's peripersonal space rather than social space. 

Defensive peripersonal space has both body part-centered and full body-centered components \cite{Serino2015}; such a representation can be learned from visuo-tactile associations on a humanoid robot and then exploited for contact prediction and whole-body avoidance behaviors \cite{Roncone2016,Nguyen_HRI_2018}. Peripersonal space representation is modulated by social context \cite{teneggi2013social}. Yet, the relationship between personal spatial zones---with origins in anthropology and social psychology---and peripersonal space, a concept from neuroscience, remains unclear despite interesting similarities in the spatial dimensions of the two concepts (see also \cite{patane2017cooperative}). The implications for HRI remain open as well. Peripersonal space---relating to safety and physical HRI---and interpersonal distance---explored in social robotics---have been separate research topics until now.

\section{Acknowledgements}
We would like to thank all the participants. Petr Ber{\'a}nek has instrumentally assisted with the experimental setup. Prof. Marek Fran{\v e}k kindly provided the Czech version on the TIPI questionnaire used in \cite{sefara2015socio}.

\bibliographystyle{IEEEtran}
\bibliography{IEEEabrv,PPSHRI}

\begin{thebibliography}{10}
\providecommand{\url}[1]{#1}
\csname url@rmstyle\endcsname
\providecommand{\newblock}{\relax}
\providecommand{\bibinfo}[2]{#2}
\providecommand\BIBentrySTDinterwordspacing{\spaceskip=0pt\relax}
\providecommand\BIBentryALTinterwordstretchfactor{4}
\providecommand\BIBentryALTinterwordspacing{\spaceskip=\fontdimen2\font plus
\BIBentryALTinterwordstretchfactor\fontdimen3\font minus
  \fontdimen4\font\relax}
\providecommand\BIBforeignlanguage[2]{{%
\expandafter\ifx\csname l@#1\endcsname\relax
\typeout{** WARNING: IEEEtran.bst: No hyphenation pattern has been}%
\typeout{** loaded for the language `#1'. Using the pattern for}%
\typeout{** the default language instead.}%
\else
\language=\csname l@#1\endcsname
\fi
#2}}

\bibitem{hall1968}
E.~T. Hall, R.~L. Birdwhistell, B.~Bock, P.~Bohannan, A.~R. Diebold~Jr,
  M.~Durbin, M.~S. Edmonson, J.~Fischer, D.~Hymes, S.~T. Kimball,
  \emph{et~al.}, ``Proxemics [and comments and replies],'' \emph{Current
  Anthropology}, vol.~9, no. 2/3, pp. 83--108, 1968.

\bibitem{lambert2004body}
D.~Lambert, \emph{Body language}.\hskip 1em plus 0.5em minus 0.4em\relax
  HarperCollins, 2004.

\bibitem{Stratton1973personal}
L.~O. Stratton, D.~J. Tekippe, and G.~L. Flick, ``Personal space and
  self-concept,'' \emph{Sociometry}, pp. 424--429, 1973.

\bibitem{mertens2014human}
A.~Mertens, C.~Brandl, I.~Blotenberg, M.~L{\"u}dtke, T.~Jacobs, C.~Br{\"o}hl,
  M.~P. Mayer, and C.~M. Schlick, ``Human-robot interaction: Testing distances
  that humans will accept between themselves and a robot approaching at
  different speeds,'' in \emph{Ambient Assisted Living}.\hskip 1em plus 0.5em
  minus 0.4em\relax Springer, 2014, pp. 269--286.

\bibitem{mutlu2008robots}
B.~Mutlu and J.~Forlizzi, ``Robots in organizations: the role of workflow,
  social, and environmental factors in human-robot interaction,'' in \emph{2008
  3rd ACM/IEEE International Conference on Human-Robot Interaction
  (HRI)}.\hskip 1em plus 0.5em minus 0.4em\relax IEEE, 2008, pp. 287--294.

\bibitem{huttenrauch2006investigating}
H.~H{\"u}ttenrauch, K.~S. Eklundh, A.~Green, and E.~A. Topp, ``Investigating
  spatial relationships in human-robot interaction,'' in \emph{2006 IEEE/RSJ
  International Conference on Intelligent Robots and Systems}.\hskip 1em plus
  0.5em minus 0.4em\relax IEEE, 2006, pp. 5052--5059.

\bibitem{rios2015proxemics}
J.~Rios-Martinez, A.~Spalanzani, and C.~Laugier, ``From proxemics theory to
  socially-aware navigation: A survey,'' \emph{International Journal of Social
  Robotics}, vol.~7, no.~2, pp. 137--153, 2015.

\bibitem{koay2017initial}
K.~L. Koay, D.~Syrdal, R.~Bormann, J.~Saunders, M.~L. Walters, and
  K.~Dautenhahn, ``Initial design, implementation and technical evaluation of a
  context-aware proxemics planner for a social robot,'' in \emph{International
  Conference on Social Robotics}.\hskip 1em plus 0.5em minus 0.4em\relax
  Springer, 2017, pp. 12--22.

\bibitem{mead2017autonomous}
R.~Mead and M.~J. Matari{\'c}, ``Autonomous human--robot proxemics: socially
  aware navigation based on interaction potential,'' \emph{Autonomous Robots},
  vol.~41, no.~5, pp. 1189--1201, 2017.

\bibitem{walters2009empirical}
M.~L. Walters, K.~Dautenhahn, R.~Te~Boekhorst, K.~L. Koay, D.~S. Syrdal, and
  C.~L. Nehaniv, ``An empirical framework for human-robot proxemics,''
  \emph{New Frontiers in Human-Robot Interaction}, 2009.

\bibitem{Syrdal2007personalized}
D.~S. Syrdal, K.~L. Koay, M.~L. Walters, and K.~Dautenhahn, ``A personalized
  robot companion? the role of individual differences on spatial preferences in
  hri scenarios,'' in \emph{RO-MAN 2007-The 16th IEEE International Symposium
  on Robot and Human Interactive Communication}.\hskip 1em plus 0.5em minus
  0.4em\relax IEEE, 2007, pp. 1143--1148.

\bibitem{Mead2016perceptual}
R.~Mead and M.~J. Matari{\'c}, ``Perceptual models of human-robot proxemics,''
  in \emph{Experimental Robotics}.\hskip 1em plus 0.5em minus 0.4em\relax
  Springer, 2016, pp. 261--276.

\bibitem{mumm2011human}
J.~Mumm and B.~Mutlu, ``Human-robot proxemics: physical and psychological
  distancing in human-robot interaction,'' in \emph{Proceedings of the 6th
  International Conference on Human-Robot Interaction}.\hskip 1em plus 0.5em
  minus 0.4em\relax ACM, 2011, pp. 331--338.

\bibitem{takayama2009influences}
L.~Takayama and C.~Pantofaru, ``Influences on proxemic behaviors in human-robot
  interaction,'' in \emph{2009 IEEE/RSJ International Conference on Intelligent
  Robots and Systems}.\hskip 1em plus 0.5em minus 0.4em\relax IEEE, 2009, pp.
  5495--5502.

\bibitem{obaid2016stop}
M.~Obaid, E.~B. Sandoval, J.~Z{\l}otowski, E.~Moltchanova, C.~A. Basedow, and
  C.~Bartneck, ``Stop! {That} is close enough. {How} body postures influence
  human-robot proximity,'' in \emph{2016 25th IEEE International Symposium on
  Robot and Human Interactive Communication (RO-MAN)}.\hskip 1em plus 0.5em
  minus 0.4em\relax IEEE, 2016, pp. 354--361.

\bibitem{Torta2011design}
E.~Torta, R.~H. Cuijpers, J.~F. Juola, and D.~van~der Pol, ``Design of robust
  robotic proxemic behaviour,'' in \emph{International Conference on Social
  Robotics}.\hskip 1em plus 0.5em minus 0.4em\relax Springer, 2011, pp. 21--30.

\bibitem{Shiomi2018}
M.~Shiomi, K.~Shatani, T.~Minato, and H.~Ishiguro, ``How should a robot react
  before people's touch? {Modeling} a pre-touch reaction distance for a robot's
  face,'' \emph{IEEE Robotics and Automation Letters}, vol.~3, no.~4, pp.
  3773--3780, 2018.

\bibitem{gosling2003}
S.~D. Gosling, P.~J. Rentfrow, and W.~B. Swann~Jr, ``A very brief measure of
  the big-five personality domains,'' \emph{Journal of Research in
  Personality}, vol.~37, no.~6, pp. 504--528, 2003.

\bibitem{github-nao-hri-personal-zones}
\BIBentryALTinterwordspacing
2019. [Online]. Available:
  \url{https://github.com/matejhof/nao-hri-personal-zones}
\BIBentrySTDinterwordspacing

\bibitem{harrigan2005proxemics}
J.~A. Harrigan, ``Proxemics, kinesics, and gaze,'' \emph{The New Handbook of
  Methods in Nonverbal Behavior Research}, pp. 137--198, 2005.

\bibitem{fiore2013toward}
S.~M. Fiore, T.~J. Wiltshire, E.~J. Lobato, F.~G. Jentsch, W.~H. Huang, and
  B.~Axelrod, ``Toward understanding social cues and signals in human--robot
  interaction: effects of robot gaze and proxemic behavior,'' \emph{Frontiers
  in Psychology}, vol.~4, p. 859, 2013.

\bibitem{Cao2018}
Z.~Cao, G.~Hidalgo, T.~Simon, S.-E. Wei, and Y.~Sheikh, ``Open{P}ose: Realtime
  multi-person 2{D} pose estimation using {P}art {A}ffinity {F}ields,'' in
  \emph{arXiv preprint arXiv:1812.08008}, 2018.

\bibitem{sefara2015socio}
D.~{\v{S}}efara, M.~Fran{\v{e}}k, and V.~Zubr, ``Socio-psychological factors
  that influence car preference in undergraduate students: the case of the
  czech republic,'' \emph{Technological and Economic Development of Economy},
  vol.~21, no.~4, pp. 643--659, 2015.

\bibitem{bartneck2009}
C.~Bartneck, D.~Kuli{\'c}, E.~Croft, and S.~Zoghbi, ``Measurement instruments
  for the anthropomorphism, animacy, likeability, perceived intelligence, and
  perceived safety of robots,'' \emph{International journal of Social
  Robotics}, vol.~1, no.~1, pp. 71--81, 2009.

\bibitem{bailenson2001equilibrium}
J.~N. Bailenson, J.~Blascovich, A.~C. Beall, and J.~M. Loomis, ``Equilibrium
  theory revisited: Mutual gaze and personal space in virtual environments,''
  \emph{Presence: Teleoperators \& Virtual Environments}, vol.~10, no.~6, pp.
  583--598, 2001.

\bibitem{maniscalco2020ass4hr}
U.~Maniscalco, A.~Messina, and P.~Storniolo, ``Ass4hr—an artificial
  somatosensory system for a humanoid robot. the ros package,''
  \emph{SoftwareX}, vol.~11, p. 100501, 2020.

\bibitem{Nguyen_HRI_2018}
P.~D. Nguyen, M.~Hoffmann, A.~Roncone, U.~Pattacini, and G.~Metta, ``Compact
  real-time avoidance on a humanoid robot for human-robot interaction,'' in
  \emph{HRI ’18: 2018 ACM/IEEE International Conference on Human-Robot
  Interaction}.\hskip 1em plus 0.5em minus 0.4em\relax ACM, New York, NY, USA,
  2018, pp. 416--424.

\bibitem{Serino2015}
A.~Serino, J.-P. Noel, G.~Galli, E.~Canzoneri, P.~Marmaroli, H.~Lissek, and
  O.~Blanke, ``Body part-centered and full body-centered peripersonal space
  representations,'' \emph{Scientific reports}, vol.~5, p. 18603, 2015.

\bibitem{Roncone2016}
A.~Roncone, M.~Hoffmann, U.~Pattacini, L.~Fadiga, and G.~Metta, ``Peripersonal
  space and margin of safety around the body: learning tactile-visual
  associations in a humanoid robot with artificial skin,'' \emph{{PLoS} {ONE}},
  vol.~11, no.~10, p. e0163713, 2016.

\bibitem{teneggi2013social}
C.~Teneggi, E.~Canzoneri, G.~di~Pellegrino, and A.~Serino, ``Social modulation
  of peripersonal space boundaries,'' \emph{Current Biology}, vol.~23, no.~5,
  pp. 406--411, 2013.

\bibitem{patane2017cooperative}
I.~Patan{\'e}, A.~Farn{\`e}, and F.~Frassinetti, ``Cooperative tool-use reveals
  peripersonal and interpersonal spaces are dissociable,'' \emph{Cognition},
  vol. 166, pp. 13--22, 2017.

\end{thebibliography}

\end{document}